
\magnification 1200
\def\be{\begin{equation}}
\def\ee{\end{equation}}
\def\bea{\begin{eqnarray}}
\def\eea{\end{eqnarray}}
\def\ba{\begin{array}}
\def\ea{\end{array}}
\def\bce{\begin{center}}
\def\ece{\end{center}}
\def\bfi{\begin{figure}}
\def\efi{\end{figure}}


\catcode`\@=11
\def\maketitle{\par
 \begingroup
 \def\thefootnote{\fnsymbol{footnote}}
 \def\@makefnmark{\hbox
 to 0pt{$^{\@thefnmark}$\hss}}
 \if@twocolumn
 \twocolumn[\@maketitle]
 \else \newpage
 \global\@topnum\z@ \@maketitle \fi\thispagestyle{empty}\@thanks
 \endgroup
 \setcounter{footnote}{0}
 \let\maketitle\relax
 \let\@maketitle\relax
 \gdef\@thanks{}\gdef\@author{}\gdef\@title{}\let\thanks\relax}
\def\@maketitle{\newpage
 \null
 \hbox to\textwidth{\hfil\hbox{\begin{tabular}{r}\@preprint\end{tabular}}}
 \vskip 2em \begin{center}
 {\Large\bf \@title \par} \vskip 1.5em {\normalsize \lineskip .5em
\begin{tabular}[t]{c}\@author
 \end{tabular}\par}
 \end{center}
 \par
 \vskip 1.5em}
\def\preprint#1{\gdef\@preprint{#1}}
\def\abstract{\if@twocolumn
\section*{Abstract}
\else \normalsize
\begin{center}
{\large\bf Abstract\vspace{-.5em}\vspace{0pt}}
\end{center}
\quotation
\fi}
\def\endabstract{\if@twocolumn\else\endquotation\fi}
\catcode`\@=12

\line{\hfill NUB-TH-3056/92}
\line{\hfill CTP-TAMU-66/92}
\line{\hfill TIFR/TH/92-69}
\line{\hfill December 15, 1992}

\bigskip

\centerline{\bf PREDICTIONS IN SU(5) SUPERGRAVITY GRAND UNIFICATION WITH}
\centerline{\bf PROTON
STABILITY AND RELIC DENSITY CONSTRAINTS}
\bigskip

\centerline{Pran Nath$^{a)}$ and R. Arnowitt$^{b)}$}
\centerline{$^{a}$Tata Institute
of Fundamental Research, Homi Bhabha Road}
\centerline{Colaba, Bombay 400 005}
\centerline{\footnote{$^*$}{Permanent address}$^a$Department of Physics,
Northeastern University, Boston, MA 02115}
\centerline{$^b$Physics Research Division, Superconducting Super Collider
Laboratory}
\centerline{Dallas, TX 75237}
\centerline{$^{\ast b}$Center for Theoretical Physics, Department of Physics,
Texas A \& M University}
\centerline{College Station, TX 77843--4242}
\centerline{[Revised]} \bigskip

\centerline{\bf Abstract}

\baselineskip=24pt

\item{} It is shown that in the physically interesting domain of the
parameter space of SU(5) supergravity GUT, the Higgs and the Z poles
dominate the LSP annihilation. Here the naive analyses on thermal
averaging breaks down and formulae are derived which give a rigorous
treatment over the poles. These results are then used to show that there exist
significant domains in the parameter space where the constraints of proton
stability and cosmology are simultaneously satisfied. New upper limits on
light particle masses are obtained.
\vfill\eject

\baselineskip=18pt

\noindent{\bf I.~ INTRODUCTION :} Recently there have been extensive
investigations of SU(5) supergravity models$^{1-7}$ with electro--weak
symmetry broken via radiative corrections$^{8,9}$. Analyses of
Refs (6)--(7) were carried out in the framework of No--Scale models$^9$ while
the analysis of Refs (1--5) are for the standard SU(5) supergravity
case$^{10}$. In this letter we shall discuss only the standard SU(5)
Model$^{10,8}$. Here, after fixing the Z--boson mass the model depends on four
arbitrary parameters, aside from the top quark mass $m_t$, which may be chosen
to be $m_o$ (the universal scalar mass), $m_{1/2}$ (the universal gaugino
mass), $A_o$ (the cubic soft SUSY breaking parameter) and tan $\beta =
<H_2>/<H_1>$ where $<H_1>$ gives mass to the down quarks and the leptons and
$<H_2>$ gives mass to the up quarks. The analyses of Refs (1--4)
investigated the full parameter space of the theory, Ref (5) investigated the
space under one more constraint ($B_0 = A_0 - m_0$ where $B_0$ is the
quadratic soft SUSY breaking parameter) while Refs (1--3) also included in
the analysis the constraint of proton stability$^{11}$. The inclusion of proton
stability constraints were seen to lead to a number of simple mass relations
among the neutralino, chargino and gluino mass spectra$^{1-3}$.
One finds for most of the parameter space $2m_{{\tilde z}_1}\cong m_{{\tilde
z}_2}\cong m_{{\tilde W}_1}$, and $m_{{\tilde W}_1}\simeq (1/4) m_{\tilde g}$
for $\mu > 0$ and $m_{{\tilde W}_1} \simeq (1/3) m_{\tilde g}$ for $\mu < 0$.
(Here ${\tilde Z}_{1,2}$ are the two lightest neutralinos, ${\tilde W}_1$ is
the lightest chargino and ${\tilde g}$ is the gluino.)  Thus the gluino mass
(approximately) determines the light neutralino and chargino spectrum.

Remarkably the standard SU(5) supergravity model under the constraint of
proton\break stability also leads to the prediction that the lightest
neutralino is the lightest supersymmetric particle (LSP)$^{1-3}$. We
investigate here the implications on the parameter space of the constraint
that the relic density of the lightest neutralino not overclose the
universe i.e. $\Omega_{{\tilde z}_1} h^2 \leq 1$, where
$\Omega_{{\tilde z}_1} = \rho_{{\tilde z}_1}/\rho_c$, with
$\rho_{{\tilde z}_1}$ the matter density of the lightest neutralino
${\tilde Z}_1$ and $\rho_c = 3H_0^2/(8\pi G_N)$ the critical density.
Here $H_0 = h \times 100 km/(s.Mpc)$ and h is the Hubble parameter with
${1 \over 2} \leq h \leq 1$. Recently$^{12}$, it was pointed out that the
dominant annihilation of neutralinos occurs near the lightest neutral
Higgs (h) pole in the s--channel for the
domain of the parameter space satisfying $CDF, LEP$ and proton stability
constraints and the finetuning requirement that
$m_{{\tilde q}, {\tilde g}} \leq 1 TeV.$ It is known that the
expansion in powers of $v$ of the thermally averaged quantity $<\sigma
v>$ (where $\sigma$ is the spin averaged annihilation cross--section of
two neutralinos, $v$ is the relative velocity defined by $\sqrt s \cong
2m_{{\tilde z}_1} + {1 \over 4} m_{{\tilde z}_1} v^2$, and $s$ is square of the
center--of--mass energy), breaks down when $\sqrt s$ is in the vicinity of
a pole$^{13}$. In this case, a careful treatment of integration over the
pole in the annihilation channel is needed. However, the rigorous analysis
even in the non--relativistic approximation involves a double integration
over the pole (which is numerically intricate) for the quantity $J =
\int_0^{x_f}dx < \sigma v>$ needed to calculate the relic density.  ($x_f =
kT_f/m_{{\tilde z}_1}$, where $T_f$ is the freeze out temperature.)
Here we derive rigorous
formulas where the integrations over one of the variables is analytically
carried out and the remaining integration is smooth over the pole. The
analysis
here is complete and includes the direct channel Higgs and Z--poles as well as
t--channel fermion exchange diagrams. Using the rigorous analysis for $J$
outlined above we explore the full five dimensional parameter space of the
theory characterized by $m_0, m_{1/2}, A_0,$ tan $\beta$ and $m_t$ under the
combined constraints of $CDF$ and $LEP$ data, proton stability and relic
density. We show that while the parameter space is strongly constrained,
significant domains in the parameter space remain where all the constraints
mentioned above are satisfied. Also new limits on the light Higgs, the light
chargino and on the $LSP$ result.

\noindent {\bf II.~ BASIC FORMULAE :} We follow standard procedure$^{14}$
and write the equation governing the number density n at time t of the
lightest neutralino ${\tilde Z}_1$ in a Friedman--Robertson--Walker universe
with isotropic mass density in the form
$$
{df \over dx} = {m_{{\tilde z}_1} \over k^3} \left({8 \pi^3N_FG_N \over
45}\right)
   ^{-{1
\over 2}} <\sigma v> \big(f^2 - f^2_0 \big)
\eqno (1)
$$
where $f = n/T^3, x = kT/m_{{\tilde z}_1}$ ($k$ is the Boltzman constant),
$N_F$
is the number of degrees of freedom at temperature T, $G_N$ is the
Newtonian constant and $f_0 = n_0/T^3$ where $n_0$ is the number density
at thermal equilibrium.  The relic density of the $LSP$ is then given by the
following (approximate) formula$^{14}$:
$$
\rho_{{\tilde z}_1} = 4.75 \times 10^{-40} \left({T_{{\tilde z}_1} \over
T_\gamma}\right)^3 \left({T_\gamma \over 2.75^\circ K}\right)^3 N_F~^{1/2}
\left({GeV^{-2} \over J(x_f)}\right) {g \over cm^3}
\eqno (2)
$$
where $(T_{{\tilde z}_1}/T_{\gamma})^3$ is a reheating factor,
$T_{\gamma}$ is the current
temperature and $J(x_f)$ is given by $J(x_f) = \int^{x_f}_0 <\sigma v> dx$
and:
$$
<\sigma v> = \int^\infty_0 dv v^2(\sigma v) e^{- v^2/4x}
\Big/\int^\infty_0 dv v^2 e^{- v^2/4x}
\eqno (3)
$$

\noindent The freezeout temperature $T_f$ is determined by the relation$^{14}$
$$
x^{-1}_f = \ell n \left[x_f^{1 \over 2}<\sigma v> \sqrt{45} m_{{\tilde
z}_1}\Big/(4\pi^3N_F^{1 \over 2} G_N^{1 \over 2}) \right]
\eqno (4)
$$
In Eq. (4) $<\sigma v>$ is the thermally average of $\sigma v$ evaluated
at $x_f$.

\noindent {\bf III.~ INTEGRATION OVER HIGGS AND Z--POLES:} $J(x_f)$
appearing in Eq.(2) can be decomposed as $J = J_h + J_Z +
J_{sf}$ where $J_h, J_Z$ are the contributions of the
s--channel Higgs and Z poles, and $J_{sf}$ is the t--channel contribution from
the exchange of squarks and sleptons.  In the domain of physical
interest with finetuning constraints $m_{{\tilde q},{\tilde g}} \leq 1 TeV$,
only
the lightest neutral Higgs $h$ makes a significant contribution to the
cross--section. For the Higgs pole, using the non--relativistic approximation,
we write $\sigma v$ in the form

$$ (\sigma v)_h = {A_h \over
m_{{\tilde z}_1}^2} {v^2 \over \big( (v^2 - \epsilon_h)^2 + \gamma_h^2 \big)}
\eqno
(5) $$
In Eq.(5) $\epsilon_h = (m_h^2 - 4m_{{\tilde z}_1}^2)/m_{{\tilde z}_1}^2$
and $\gamma_h = m_h\Gamma_h/m_{{\tilde z}_1}^2$ where $m_h$ is the Higgs mass
and
$\Gamma_h$ is the Higgs decay width and $A_h$ is$^{15}$

$$ \eqalignno{A_h=&{1 \over 8\pi}
\left({g_2 \over 2M_W} {\sin \alpha \over \cos \beta}
\right)^2 {g^2_2 \over \cos^2 \theta_W} \big(n_{11} \cos\theta_W -
n_{12}\sin\theta_W \big)^2 \cr & \big(n_{13} \sin\alpha + n_{14} \cos\alpha
\big)^2 \sum_i C_im_{fi}^2 \left(1 -{m_{fi}^2 \over m_{{\tilde
z}_1}^2}\right)^{3 \over 2}~. & (6) \cr} $$
Here $\sin 2\alpha = - (m_A^2 +
m_Z^2)(m_H^2 - m_h^2)^{-1} \sin 2\beta$ and where $m_A$ is the mass of the
CP--odd Higgs and $m_H$ is the mass of the CP--even heavy neutral Higgs. $C_i$
is
a color factor which is (3,1) for (quarks, leptons) and $n_{1i}$ are
components of ${\tilde Z}_1$ eigen--vector in the basis defined in Ellis et al
in Ref (14). Using Eq.(5) one can carry out the $x$--integration in $J_h
(x_f)$ and get

$$ J_h(x_f) = {A_h \over 2\sqrt{2} m_{{\tilde
z}_1}^2} \left[I_{1h} + {\epsilon_h \over \gamma_h} I_{2h} \right] \eqno (7)
$$
where

$$ I_{1h} = {1 \over 2} \int^\infty_0 d\xi \xi^{-{1 \over 2}} e^{-
\xi} \ell n \left[{(4 \xi x_f - \epsilon_h)^2 + \gamma_h^2 \over \epsilon_h^2 +
\gamma_h^2} \right] \eqno (8) $$
$$ I_{2h} = \int^\infty_0 d\xi \xi^{-{1 \over 2}}
e^{-\xi} \left[\tan^{-1} \left({4
    \xi
x_f - \epsilon_h \over \gamma_h}\right) + \tan^{-1}
\left({\epsilon_h \over \gamma_h}
\right) \right]
\eqno (9)
$$
A similar analysis can be carried out for the Z--Pole and here one finds

$$
J_Z = {1 \over 2 \sqrt{\pi} m_{{\tilde z}_1}^4} \left[A_Z {I_{1Z} \over
\epsilon_Z} + {\epsilon_Z \over \gamma_Z} B_Z \big(I_{1Z} + I_{2Z} \big)
\right]
\eqno (10)
$$
where $I_{1Z}$ and $I_{2Z}$ are defined anologously to $I_{1h}$ and $I_{2h}$
with $m_h, \Gamma_h$ replaced by $M_Z, \Gamma_Z$. In Eq.(10) $A_Z$ is
given by

$$A_Z={\pi \over 8} {\alpha_2^2 \over \cos^4\theta_W} (n_{13}^2 -
n_{14}^2)^2 \left (1-{4m_{{\tilde z}_1}^2 \over M_z^2}\right )^2$$
$$\times\left [ 3m_b^2 \Big(1 - {m_b^2 \over m_{{\tilde z}_1}^2} \Big)^{1 \over
2}
+ m_{\tau}^2 \Big(1 - {m^2_{\tau} \over m_{{\tilde z}_1}^2} \Big)^{1 \over
2}+3m^2_c\left ( 1 - {m^2_c\over m^2_{{\tilde z}_1}}\right )^{1\over 2}
\right ]\eqno (11)
$$
where we have retained only the dominant $b$, $c$
and $\tau$--contributions, while
$B_Z$ (in the zero--fermion mass approximation) is

$$
B_Z = {\pi \over 6} {\alpha^2_2 \over \cos^4 \theta_W} m^2_{{\tilde z}_1}
(n_{13}^2 - n_{14}^2)^2 \left[{21 \over 2} + {80 \over 3} \sin^4 \theta_W - 20
\sin^2 \theta_W \right]
{}~.\eqno (12)
$$
In the vicinity of the Higgs (or Z--Pole), $J_{sf}$ is typically much
smaller than $J_h$ (or $J_Z$) and thus we shall use the
conventional approximation$^{14-15}$ of $ax_f + {1 \over 2} bx_f^2$ in
computing $J_{sf}$.

\noindent {\bf IV.~ ANALYSIS AND RESULTS :} We begin by exhibiting the
result that the computation of $J_{approx}$ using power expansion in
$v^2$ on $<\sigma v>$ is a poor approximation to the full analysis of J
where rigorous thermal averaging on the Higgs and Z--Poles is carried out.
The ratio of $\Omega_{approx}/\Omega = J/J_{approx}$ is exhibited in Fig. 1.
The results of Fig. 1 show that $\Omega_{approx}$ can be
inaccurate by up to 3 orders
of magnitude and show a total breakdown of the approximate result near the
Higgs pole or $Z$ pole.

To proceed further we must include proton stability constraints. In
supergravity SU(5), the dominant proton decay proceeds via dimension five
operators and involves the Higgs color
triplet exchange. The most dominant decay
mode is $p \rightarrow \overline \nu K^+$, and
proton stability may be conveniently characterized by the value of the
dressing loop function B that enters in $p \rightarrow \overline \nu K^+$
decay and is defined in Ref 1. The current Kamiokande bound of$^{16}$ $\tau (p
\rightarrow \overline \nu K^+) > 1 \times 10^{32} yr$ translates to
a bound on B of$^{17}$
$$
B < 105 \left({M_{H_3} \over M_G} \right) GeV^{-1}
\eqno (13)
$$
where $M_{H_3}$ is the Higgs triplet mass and $M_G$ is the GUT mass. We
also note that the simplest GUT sector in SU(5)$^{18}$ leads to the
relation $M_{H_3} / M_V = (\alpha_\lambda / \alpha_G)^{1 \over 2}$
between the Higgs triplet mass $M_{H_3}$ and the
massive vector boson mass $M_V$.
[Here
$\alpha_\lambda = \lambda^2_2/4\pi$ and $\lambda_2$ enters the
GUT superpotential via the term $\lambda_2 H_1(\Sigma + 3 M){\bar H}_2$
where
$H_1, {\bar H}_2$ are the 5, $\bar 5$ and $\Sigma$
is the 24--plet representation of
SU(5).] An upper limit on the Higgs triplet mass emerges if one
assumes that the Yukawa couplings be perturbative at the GUT scale.
Estimates on $M_{H_3}$ that lead to perturbative $\lambda_2$ lie in the
range $M_{H_3} < 3M_G^{1,2}$ to $M_{H_3} < 10 M_G^{19}$. Here as a guideline we
shall use a benchmark limit of $M_{H_3} < 6 M_G$.

We discuss now the result of the analysis. We start at the GUT scale with
SU(5) supergravity boundary conditions and use the renormalization group
equations to evolve masses and coupling constants to low energy where a
radiative breaking of the electro--weak symmetry is achieved. Solutions
are subjected to constraints of the $CDF$ and $LEP$ data which give lower
limits on the SUSY mass spectra, the proton decay constraint of Eq.(13) with
$M_{H_3} < 6 M_G$, and the relic density constraints discussed in secs
I--III. We shall also impose the fine tuning condition $m_{{\tilde q},
{\tilde g}} < 1 TeV$. The analysis shows that there
exists significant domains in the parameter space for both $\mu > 0$
and $\mu < 0$ ($\mu$ is the Higgs
mixing parameter which enters the superpotential via the term $\mu H_1
H_2$) where all
the desired constraints are satisfied.  The allowed parameter space is found
to be larger for the case $\mu > 0$.

We discuss the $\mu > 0$ case now in greater detail. Fig. 2 exhibits the
allowed domain consistent with proton stability and relic density for
the case $m_0 = 700 GeV$ as a
function of $A_t$ (where $A_t$ is the t-quark Polonyi constant at the
electro--weak scale) when tan $\beta$ and $m_{1/2}$ are
varied over the allowed range. Fig. (3) exhibits the allowed domain as a
function of $\alpha_H (\tan \alpha_H \equiv \cot \beta)$ at $A_t = 0$ when
$m_0$ and $m_{1/2}$ are varied over the allowed range of values. In each
of the two cases one finds that the domain of the parameter space
consistent with $CDF, LEP$ data, proton stability and relic density is
quite substantial even at the lower bound of $B < 300$ GeV$^{-1}$ ($M_{H_3} =
3 M_G$).

New upper limits on the Higgs mass and on the chargino mass also emerge.
One finds that $m_h \leq 105$ GeV and $m_{{\tilde W}_1} < 100$ GeV for
$B < 600$ GeV$^{-1}$.
Thus the chargino should be seen at $LEP2$ while for much of the allowed
parameter space,
the light CP even Higgs should also be seen.  The lightest
neutralino mass has an upper limit of $\l 50$ GeV
and the maximum t-quark mass is $\simeq 165$ GeV.  These bounds are
lower than the ones given in Ref. 2 where no relic density constraint
was imposed.  The $h$, ${\tilde W}_1$ and ${\tilde Z}_1$ mass bounds also
decrease if one lowers the bound on $B$.

\noindent {\bf IV.~ CONCLUSION :} It is shown that in the physically
interesting domain of the parameter space of the standard SU(5)
supergravity, annihilation of the relic neutralinos is dominated by the
light Higgs and the Z poles. Analysis is given which treats the thermal
averaging over the poles rigorously. Previous approximate analyses are found to
be inaccurate by several orders of magnitude near the Higgs pole and also
significantly inacurate near the Z pole. It is found that
for both $\mu > 0$ and $\mu < 0$
significant
domains of the parameter space exists where all the desired constraints are
satisfied.

\noindent {\bf ACKNOWLEDGEMENTS :} This research was supported in part by
NSF Grant Nos. PHY--916593 and PHY--917809. One of us (P.N.) thanks TIFR
where part of this work was done for the hospitality accorded him.
We wish to thank Jorge Lopez and Kajia Yuan for pointing out an incorrect
reheating factor (taken from Ellis et al., Ref. 14) in our earlier version of
this paper.  The correction leads to a much wider region where relic density
constraints are satisfied.
\bigskip
\bigskip
\smallskip

\centerline {\bf Figure Captions}
\bigskip
Fig. 1 : $\Omega_{approx}/\Omega$ as a function of $m_{gluino}$ for top
masses of 110 GeV (dashed curve), 125 GeV (solid curve) and 140 GeV
(dotted curve), showing
massive breakdown of the approximation near the Higgs and Z poles.  The poles
occur close to where $\Omega_{approx}/\Omega$ decreases sharply.  Note that
$\Omega_{approx}$ is least accurate in the region prior to the poles, which is
also where $\Omega h^2<1$.
\bigskip \bigskip

Fig. 2 : Allowed domains in the $B-A_t$ plane for top mass of 110 GeV (dashed
curves), 125 GeV (solid curves) and 140 GeV (dotted curves) when $m_0$ = 700
GeV.
The domain allowed by relic density constraints is the region between the
upper and lower curves.
The domain allowed by proton stability lies below the solid horizontal
line when $M_{H_3} < 6 M_G$.  The gap in the central region for $m_t = 110$
GeV is due to the requirement that $m_h > 60$ GeV.
\bigskip \bigskip

Fig. 3 : Allowed domains in the $B-\alpha_H$ (tan $\alpha_H \equiv
ctn~\beta$) plane for top quark masses of 110 GeV (dashed curves), 125 GeV
(solid curves), 140 GeV (dotted curves) and 160 GeV (dot--dash curves), when
$A_t = 0$. The domain allowed by
relic density constraints and proton stability is as in Fig. 2.

\baselineskip=.5cm

\noindent {\bf REFERENCES :}
\smallskip
\item{1.} R. Arnowitt and P. Nath, Phys. Rev. Lett. \underbar {69}, 725 (1992).

\item{2.} P. Nath and R. Arnowitt, Phys. Lett. \underbar {B289}, 368 (1992).

\item{3.} J. Lopez, H. Pois, D.V. Nanopoulos and K. Yuan,
CERN--TH--6628/92--CTP--TAMU--61/92-ACT-19/92.

\item{4.} G.G. Ross and R.G. Roberts, Nucl. Phys. \underbar {B377}, 571 (1992).

\item{5.} M. Drees and M.M. Nojiri, Nucl. Phys. \underbar {B369}, 54 (1992).

\item{6.} K. Inoue, M. Kawasaki, M. Yamaguchi and T. Yanagida, Phys. Rev.
\underbar {D45}, 387 (1992); S. Kelley, J. Lopez, H. Pois, D.V. Nanopoulos and
K. Yuan, Phys. Lett. \underbar {B273}, 423 (1991).

\item{7.} P. Nath and R. Arnowitt, Phys. Lett. \underbar {B287}, 89 (1992).

\item{8.} For a review see H.P. Nilles, Phys. Rep. \underbar {110}, 1 (1984).

\item{9.} For a review see A.B. Lahanas and D.V. Nanopoulos, Phys. Rep.
\underbar {145}, 1 (1987).

\item{10.} A.H. Chamseddine, R. Arnowitt and P. Nath, Phys. Rev. Lett.
\underbar
{49}, 970 (1982).

\item{11.} J. Ellis, D.V. Nanopoulos and S. Rudaz, Nucl. Phys. \underbar
{B202}, 43 (1982); R. Arnowitt, A.H. Chamseddine and P. Nath, Phys. Lett.
\underbar {156B}, 215 (1985);
P. Nath, A. H. Chamseddine, and R. Arnowitt, Phys. Rev. \underbar{D32}, 2348
(1985) and references quoted there in.

\item{12.} R. Arnowitt and P. Nath, Phys. Lett. \underbar{B299}, 58 (1993).

\item{13.} K. Griest and D. Seckel, Phys. Rev. \underbar {D43}, 3191 (1991): P.
Gondolo and G. Gelmini, Nucl. Phys. \underbar {B360}, 145 (1991).

\item{14.} B.W. Lee and S. Weinberg, Phys. Rev. Lett. \underbar {39}, 165
(1977); H. Goldberg, Phys. Rev. Lett. \underbar {50}, 1419 (1983); J. Ellis,
J.S. Hagelin, D.V. Nanopoulos, K. Olive and M. Srednicki, Nucl. Phys.
\underbar {B238}, 453 (1984).

\item{15.} J. Lopez, D.V. Nanopoulos and K. Yuan, Nucl. Phys. \underbar
{B370}, 445 (1992); M. Drees and M.M. Nojiri, DESY92--101.

\item{16.} Particle Data Group, Phys. Rev. \underbar{D45}, Part 2 (1992).

\item{17.} Eq. (13) differs slightly from the result quoted in Ref. 1 as we
use here more recent values of $\alpha_i (M_Z)$.

\item{18.} E. Witten, Nucl. Phys. \underbar {B177}, 477 (1981); S. Dimopoulos
and H. Georgi, Nucl. Phys. \underbar {B193}, 150 (1981); N. Sakai, Z. Phys.
\underbar {C11}, 153 (1981).

\item{19.} H. Hisano, H. Murayama and T. Yanagida, Tohuku University preprint
TU--400--July, 1992.

\bye